\documentclass{elsart}

\usepackage{epsf}
\input epsf

\newcommand{\mpi}{Max-Planck-Institut f{\"u}r Kernphysik}
\newcommand{\CE}{\v{C}erenkov}

\begin{document}
\begin{frontmatter}

% -- title -- %
\title{Particle Identification by Multifractal Parameters in $\gamma$-Astronomy
with the HEGRA-{\CE}-Telescopes}

\collab{HEGRA Collaboration}

\author{B. M. Sch{\"a}fer},
\author{W. Hofmann}, 
\author{H. Lampeitl}, 
\author{M. Hemberger}

\address{\mpi, P.O. Box 103980, D-69029 Heidelberg, Germany}

% +++++++++++++++++++++++++++++++++++++++++++++++++++++++++++++++++++++++++++ %
% +++++                          ABSTRACT                               +++++ %
% +++++++++++++++++++++++++++++++++++++++++++++++++++++++++++++++++++++++++++ %
\begin{abstract}
{\CE} images of air showers can also be classified using 
multifractal and wavelet parameters, as compared to the conventional
Hillas image parameters. This new technique was applied to the images
recorded by the cameras of the stereoscopic imaging air {\CE}-telescopes
operated by the HEGRA collaboration. 
With respect to the identification of particles,
the performance of multifractal and
wavelet parameters was examined using a data 
sample from the observation of the active galaxy Mkn~501 that showed a
high $\gamma$-ray flux. The multifractal
parameters were also combined
with the Hillas parameters using a neural network approach in
order to further improve the $\gamma$/hadron-separation.
\end{abstract}

% -- Keywords -- %

\end{frontmatter}

% +++++++++++++++++++++++++++++++++++++++++++++++++++++++++++++++++++++++++++ %
% +++++                          INTRODUCTION                           +++++ %
% +++++++++++++++++++++++++++++++++++++++++++++++++++++++++++++++++++++++++++ %
\section{Introduction}

Over the last decade, imaging air {\CE}-telescopes have emerged as the most
powerful tools for $\gamma$-ray astronomy in the TeV energy range. The imaging
of air showers allows a separation of $\gamma$-ray induced air showers and
hadron-induced air showers, and therefore a strong reduction of the
cosmic-ray background (\cite{CTreview1,CTreview2}). The performance of such
instruments can be boosted further by combining several {\CE}-telescopes
to form a stereoscopic system, providing several simultaneous views of
an air shower. For example, the HEGRA system of {\CE}-telescopes
(\cite{PerformancePaper,Mrk501time,Mrk501spectrum}) achieves an angular 
resolution of $0.1^\circ$ for individual photons, and an energy resolution
of better than 20\%.

In the traditional approach, 
the distribution of the {\CE} light in the camera is characterized
by an ellipse,
whose semi-axes are derived by the calculation of the second moments of the
amplitude distribution after a suitable procedure to clean the image
of singular high amplitudes at a large distance from the actual projection of
the shower (\cite{HIL85,FEG97}). The width of this ellipse very 
efficiently distinguishes between electromagnetic and hadronic showers. 
To remove the dependence of the image width on zenith angle, 
total amplitude and 
distance of the shower core to the telescope, one can use
the {\it scaled width} (\cite{mscwid}) where the {\it width} parameter is 
divided by the Monte-Carlo expectation value for $\gamma$-rays. 
For a system of {\CE}-telescopes, 
the {\it scaled width} is averaged over all triggered telescopes. 
The resulting {\it mean scaled width} 
of the image ellipses peaks at a value of 1.0 for 
$\gamma$-originated showers and shows excellent performance with 
respect to separating $\gamma$-rays from the hadronic background
(\cite{mscwid}). Thus, cuts on 
{\it mean scaled width} enhance the significance of a 
$\gamma$-signal strongly.\\

In an alternative approach, one may visualize the cascading process in which a 
shower develops as a series of self-similar processes and hence one expects 
that the distribution of {\CE} light exhibits self-similar structures, as was
pointed out in \cite{HAU99}. The possibility exists that in a shower of reasonable size 
the superposition of the {\CE} light from all sub-showers results in a smooth image in 
which the these structure has been averaged out, especially when taking account of the fact
that the {\CE} light is subjected to scattering processes. Therefore, the {\CE} light 
distribution recorded by the camera does not necessarily have to have fractal 
properties.

Multifractal analysis aims at describing this 
self-similarity quantitatively. The scope of the article is to give a brief 
introduction to multifractal and wavelet analysis (section 2), to describe
how these methods were applied to the images recorded by the HEGRA-cameras, 
and to 
investigate their properties in determining the nature of the primary particle 
by comparing their performance on a Mkn~501 data sample 
with the approach by Hillas (section 3). 
An attempt was made to combine both Hillas parameters 
and multifractal parameters using neural networks (section 4).

% +++++++++++++++++++++++++++++++++++++++++++++++++++++++++++++++++++++++++++ %
% +++++                    MULTIFRACTAL ANALYSIS                       ++++++ %
% +++++++++++++++++++++++++++++++++++++++++++++++++++++++++++++++++++++++++++ %
\section{Multifractal Analysis}

\subsection{Multifractals}

The quantities derived in multifractal analysis describe the behaviour of
patterns with respect to scale transformations. 
We assume that the pattern -- i.e., the {\CE} image -- is described as
a density $A(\vec{x}\prime)$ defined on a two-dimensional grid, 
corresponding to the 
amplitudes measured in the photomultiplier pixels of a camera.
In a first step, the distribution is normalized such that
$\sum_{\left\{\vec{x}\prime\right\}}A(\vec{x}\prime) = 1$.
Starting point of the analysis is the
calculation of a multifractal moment $Z_q(\ell)$ of order $q$, defined on a
length scale $\ell$ (\cite{FED88})
\begin{equation}
Z_q(\ell) = \sum_{\left\{\vec{x}\right\}}p(\vec{x},\ell)^q
\qquad\mbox{with}\qquad p(\vec{x},\ell) = 
\sum_{\vec{x}\prime\mbox{ }\in\mbox{ }\parbox{2.2cm}{\baselineskip 0.27cm 
{\mbox{\scriptsize box of size $\ell$}\\ 
 \mbox{\scriptsize centered at $\vec{x}$}}}}
A(\vec{x}\prime).
\end{equation}
Here, the image is divided up into a set of boxes of characteristic
scale $\ell$; the non-overlapping boxes have identical shapes and
cover the image completely.
The amplitudes 
$A(\vec{x}\prime)$ within a box of size $\ell$ at a lattice site $\vec{x}$
are summed up, yielding $p(\vec{x},\ell)$. These so-called box-amplitudes
are exponentiated with a number $q$, that assumes integer values, 
and successively the summation over a set boxes 
is formed. \\

With varying lengthscale $\ell$ one expects for a genuine fractal distribution 
a behaviour like
\begin{equation}
Z_q(\ell)\sim\ell^{\tau(q)}\mbox{,}\label{loglog}
\end{equation}
from which the first multifractal parameter, the mass exponent $\tau(q)$, may 
be derived as the slope of the line through the points $(Z_q(\ell)|\ell)$ in a 
double-logarithmic plot. The order $q$ of the moment defines how 
structures that are less bright and prominent contribute to the value of 
$Z_q(\ell)$. The singularity exponent $\alpha(q)$, defined as
\begin{equation}
\alpha(q) = \frac{d}{dq} \tau(q)\mbox{,}
\end{equation}
describes how $\tau(q)$ changes with changing order $q$.

\subsection{Wavelets}

Wavelet analysis is closely related to multifractal analysis: it describes the
scaling property of differences in amplitude in adjacent boxes rather than
the amplitudes themselves, i.e. the derivative of the amplitude distribution on
different baselines. Alternatively one may think of wavelet analysis as a
Fourier expansion with a discrete set of basis functions that are generated by
scaling from a so-called mother wavelet (\cite{BAC93}).\\

In analogy with multifractal analysis (\cite{KAN95}), one defines a 
wavelet moment whose behaviour under scale transformation is determined by 
an exponent $\beta$. For a one-dimensional density, the wavelet moment
is given by
\begin{equation}
W_q(\ell) = \sum_{\left\{\vec{x}\right\}}
\left|p(\vec{x},\ell)-p(\vec{x}+\ell\cdot\vec{e}_x,\ell)\right|^q
\Longrightarrow W_q(\ell)\sim\ell^{\beta(q)}\mbox{.}
\end{equation}
The relevant quantity for the derivation of a wavelet moment is the differene in
integrated amplitudes between one box of linear extend $\ell$ and the neighboring box in 
the direction indicated by the unit vector $\vec{e}_x$. This corresponds to the 
convolution of the amplitude distribution with a wavelet (in this case the simplest 
so-called Haar wavelet), anlyzing the distribution with at spatial resolution $\ell$. 
The sign of the wavelet function is indicated pictorially by $(+-)$.\\

In order to characterize a two-dimensional distribution on a scale $\ell$ by its wavelet
moments, a set of three base wavelets 
${+\,+\choose-\,-}$, ${+\,-\choose+\,-}$, ${+\,-\choose-\,+}$ is required (\cite{KAN95}), 
each yielding a wavelet moment and a corresponding exponent.

\subsection{Application to the Shower Images}

The HEGRA-{\CE}-telescopes (\cite{HEGRA})
are equipped with cameras (\cite{camera}) consisting
of 271 pixels of $0.25^{\circ}$ diameter arranged on a hexagonal lattice. 
The telescopes are triggered by a coincidence
of two pixels with at least 6 to 8 photoelectrons each. Images
contain about 100 photoelectrons per TeV of $\gamma$-ray energy.
Therefore, images near the trigger threshold (0.5 TeV) contain about 50
photoelectrons, typical images about 100 photoelectrons, and 
the largest gamma-ray images about 2000-3000 photoelectrons.
Night-sky background light as well as electronics noise and
ADC quantization cause a typical noise of 1 photoelectron rms
in each pixel. For the normal Hillas-type analysis, only pixels
with signals of at least than 6 photoelectrons are used, or pixels
with at least 3 photoelectrons provided that they are adjacent
to a pixel with at least 6 photoelectrons.

\begin{figure}[htb]
\begin{center}
\begin{tabular}{cc}
 \mbox{
    \epsfxsize5cm
    \epsffile{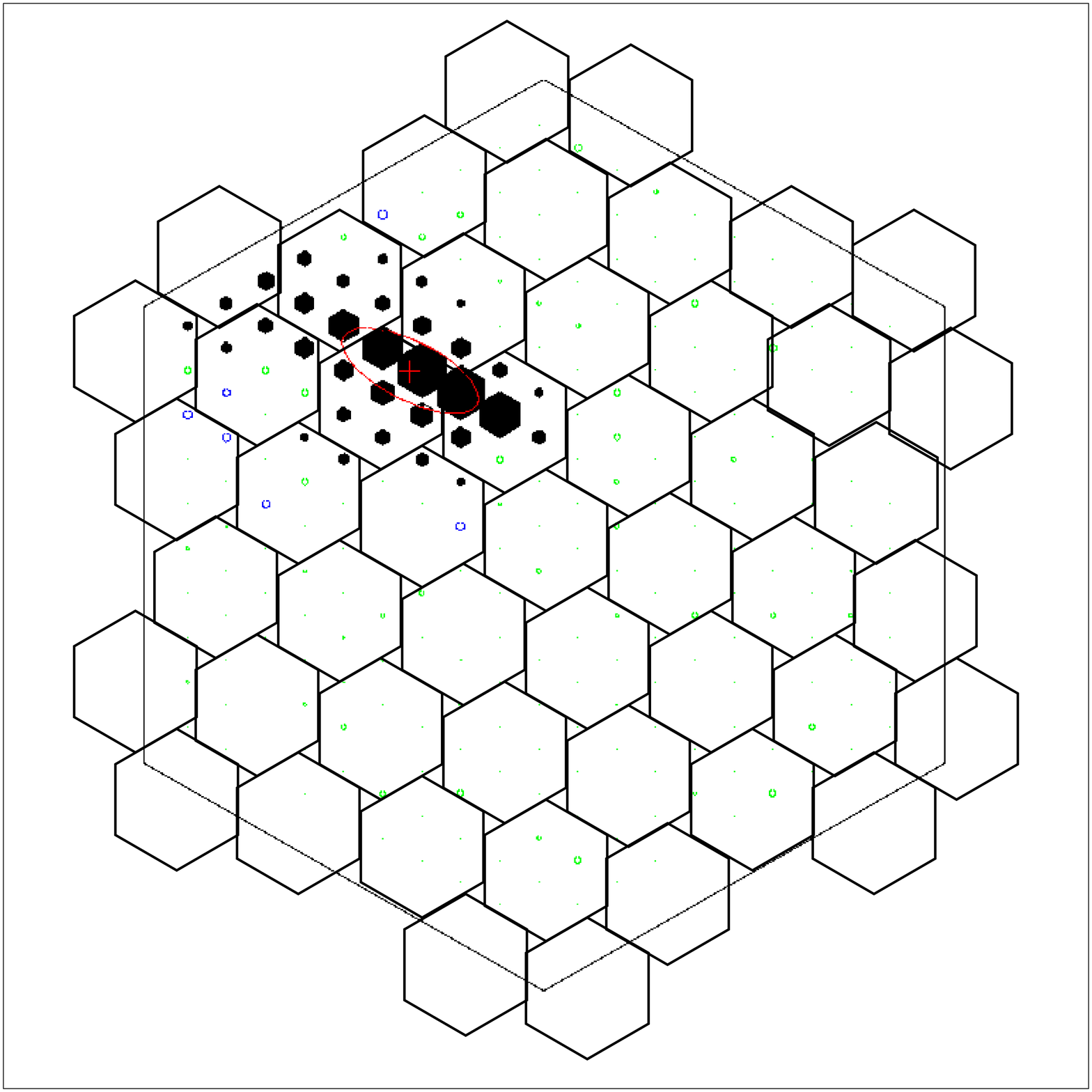}
      }
 & 
 \mbox{
    \epsfxsize5cm
    \epsffile{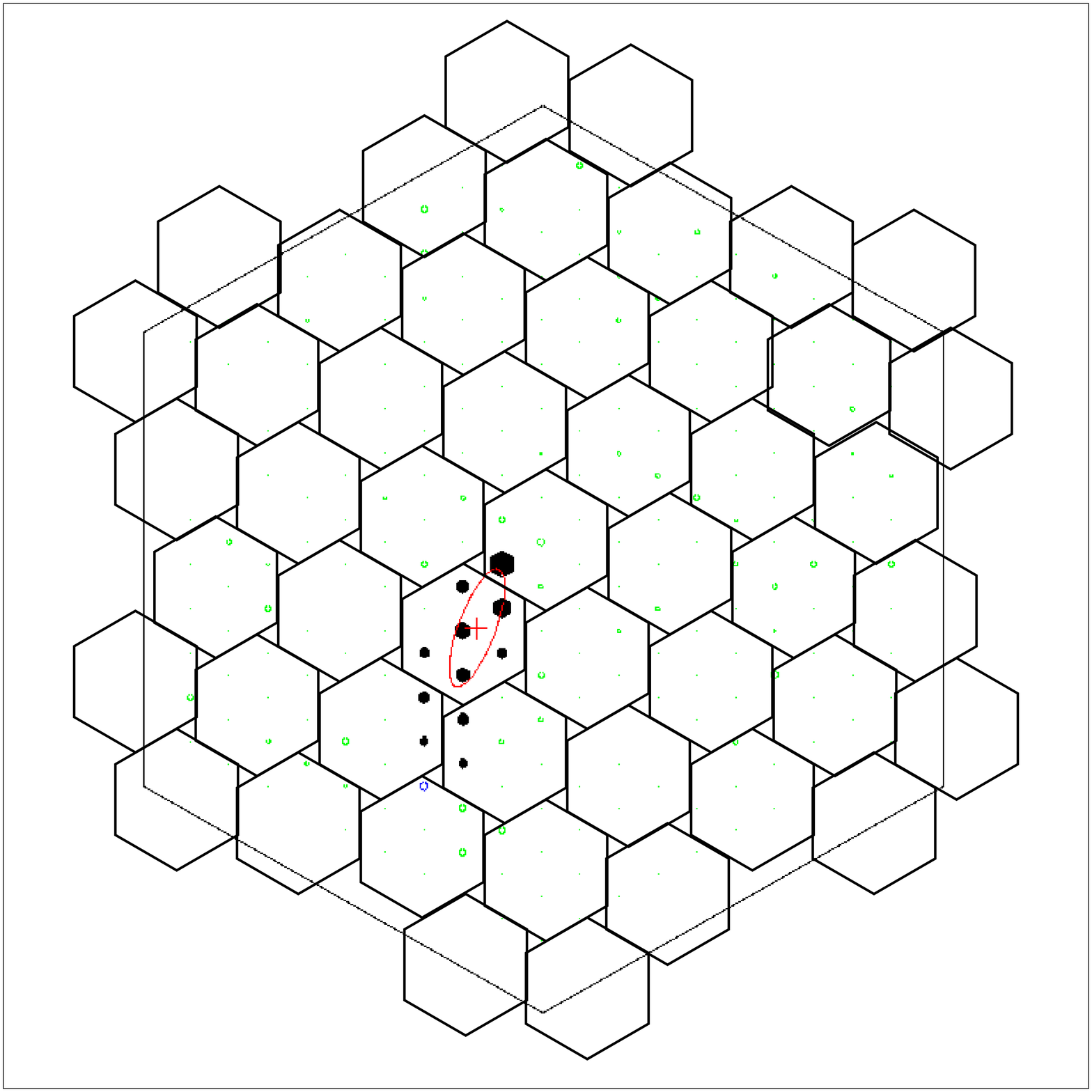}
      }
 \\
 \mbox{
    \epsfxsize5cm 
    \epsffile{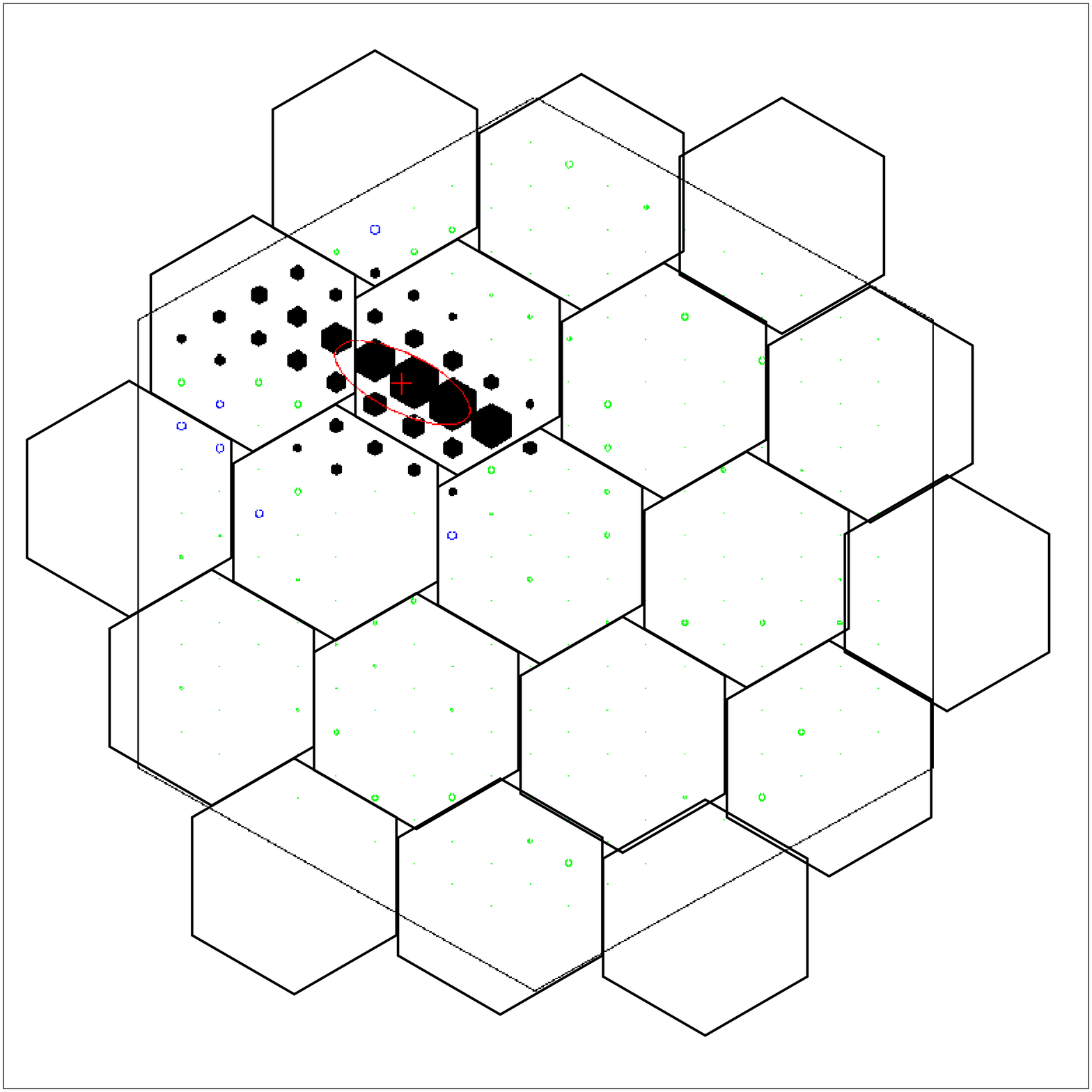}
      }
 &
  \mbox{
    \epsfxsize5cm
    \epsffile{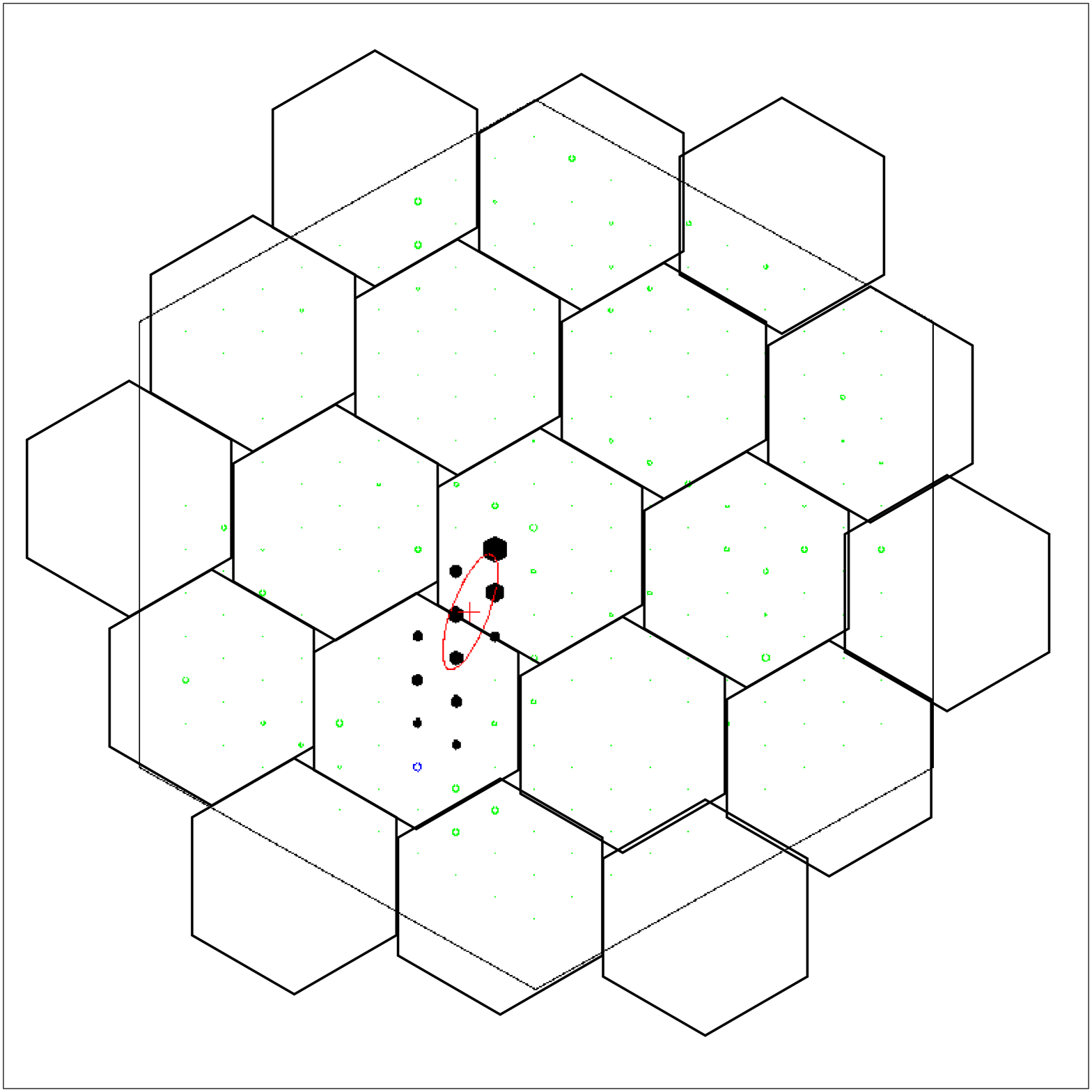}
      }
 \\ 
 \mbox{
    \epsfxsize5cm 
    \epsffile{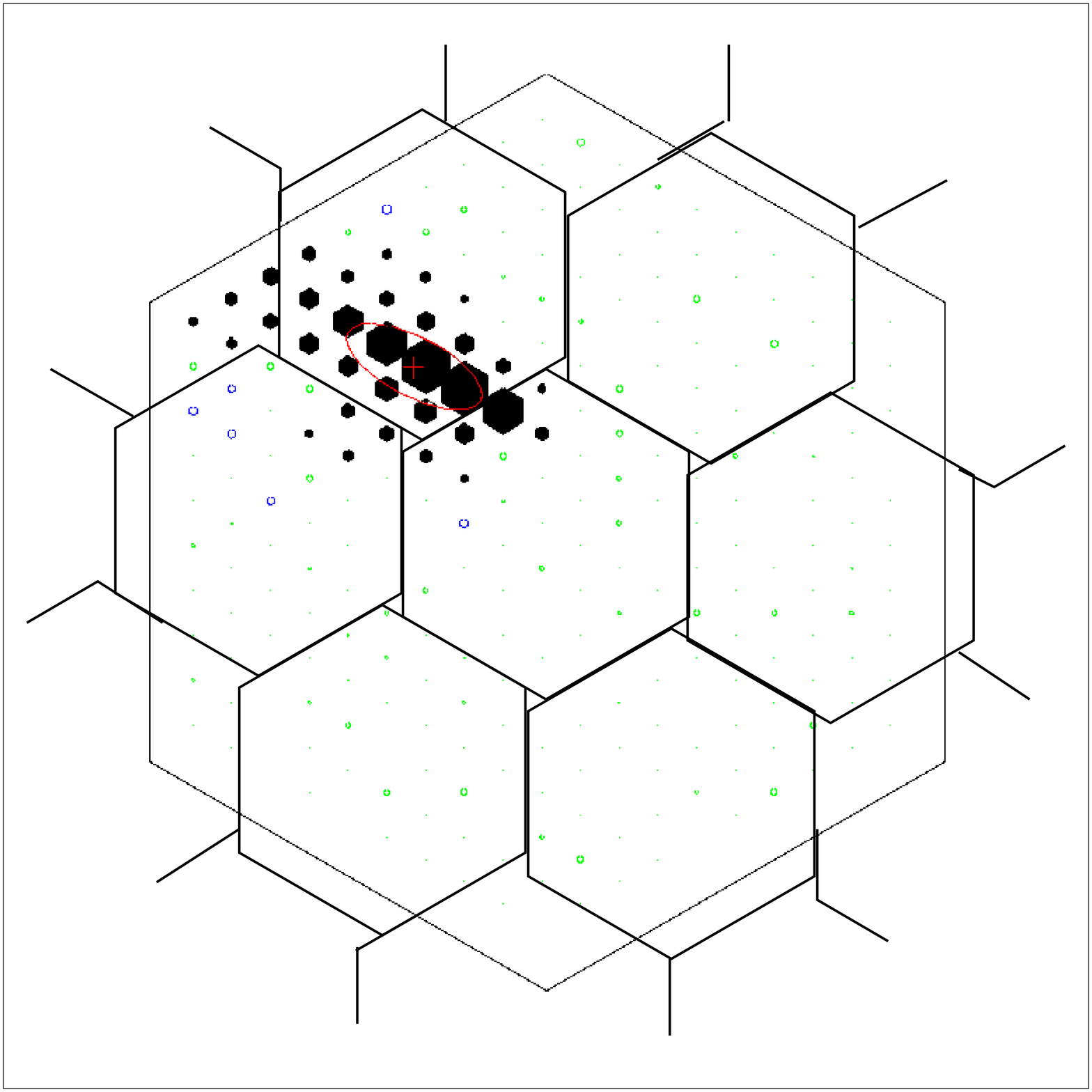}
      }
 & 
 \mbox{
    \epsfxsize5cm
    \epsffile{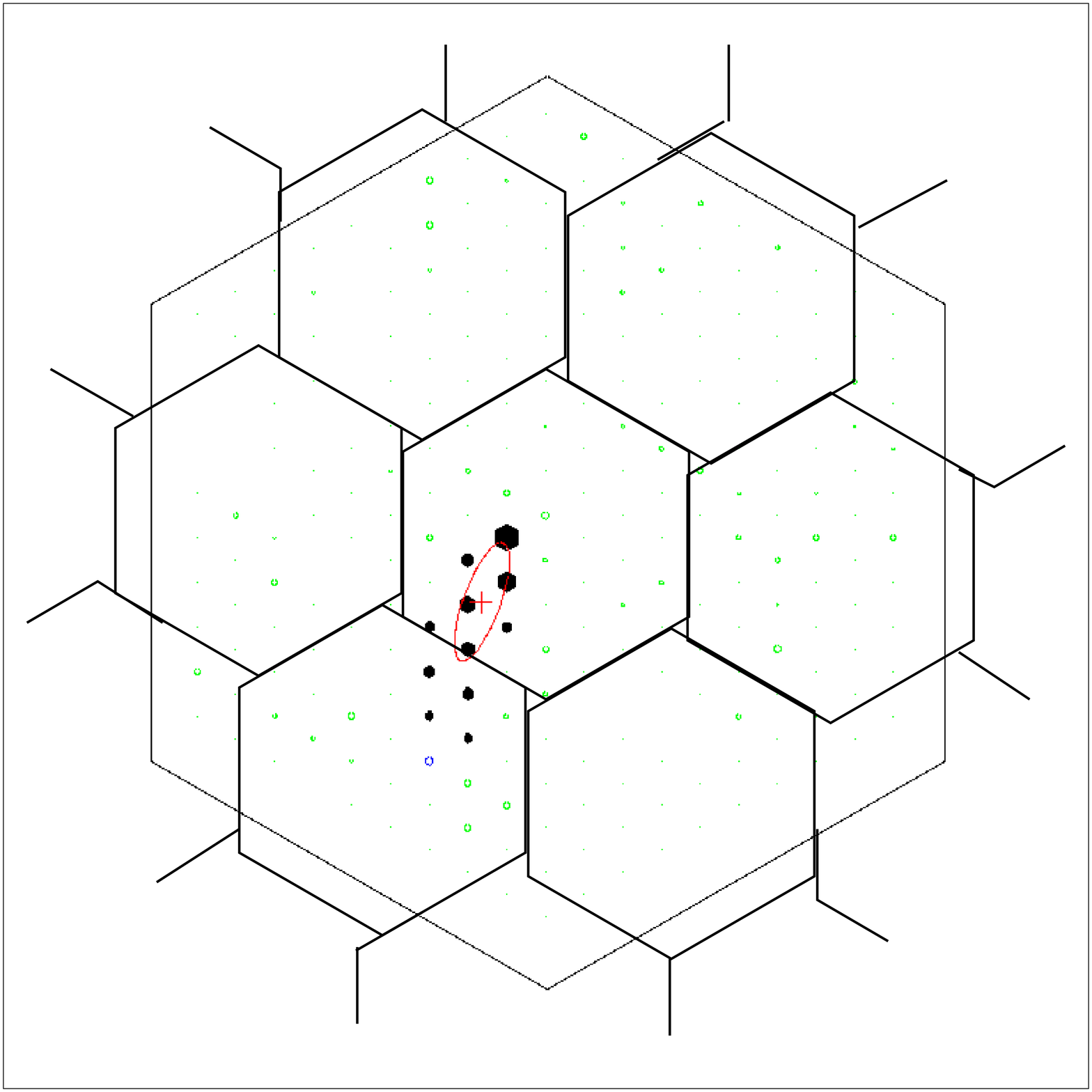}
      }
\end{tabular}
\end{center}
\caption{Definition of the cells of pixels for $\ell=2$ (top row), $\ell=3$
(middle row) and $\ell=4$ (bottom row), superimposed to a shower 
image containing about 2000 photoelectrons (left column) and about 
100 photoelectrons (right column). The cells corresponding to $\ell=1$ are the
individual pixels. The ampitude in a given pixel is indicated by the block area.
The thin line indicates the border of the camera.}
\label{pixgroup}
\end{figure}

The hexagonal lattice of the cameras 
complicates the fractal analysis slightly. Unlike in the
case of a square lattice (\cite{HAU99}), it is not advisable to form ``boxes'' 
of 1, $4=2\times2$, $9=3\times3$, ... pixels, since the resulting ``boxes'' 
have a preferred axis.
Instead, hexagonal cells consisting of 1, 7, 19 and 37 pixels were 
defined; the size of one cell is described by the extent of the hexagon along 
one side, thus $\ell$ assumes values of $\ell = \mbox{1, 2, 3 and 4}$, as it is 
shown in Figure~\ref{pixgroup}. 

Another complication arises from the noise contributions. There is a 
finite probability to have a very small signal in a pixel, which would
make moments with negative index $q$ diverge. Since the amplitude
distribution is more or less continuous - the peaks corresponding to
0, 1, 2, ... photoelectrons are not resolved individually - such pixels
could only be removed by a more or less arbitrary cut. Instead, it was
decided to limit the analysis to positive values of $q$, 
in the range from 2 to 8. The values $q=0$ and $q=1$ yield trivial 
results; $q=0$ simply counts the numbers of cells and the moment
for $q=1$ simply reflects the normalization of the amplitudes. 

% +++++++++++++++++++++++++++++++++++++++++++++++++++++++++++++++++++++++++++ %
% +++++                            RESULTS                              +++++ %
% +++++++++++++++++++++++++++++++++++++++++++++++++++++++++++++++++++++++++++ %
\section{Results}

The analysis methods described above were applied to a subset of the data 
resulting from the observation of the active galaxy Mkn~501 in a high 
state of activity in the course of the year 1997 
(\cite{Mrk501time}, \cite{Mrk501spectrum}).  
The data were taken at zenith angles between $10^{\circ}$ and $26^{\circ}$.
During these observations, the source was displaced by $0.5^\circ$ in declination
with respect to the optical axis of the telescopes; a region displaced
by the same amount in the opposite region served to evaluate backgrounds
under the signal. The sign of the displacement varied between 20-minute
runs, in order minimize systematic effects. The distance of $1^\circ$
between the signal region and the background region is large enough
that events reconstructed stereoscopically can be assigned unambiguosly.

After a preselection keeping only events with shower axes within
$0.5^\circ$ from the source or from the center of the background region,
the data set contained approximately 29.000 events originating from the region 
centered on the source position opposed to 25.000 hadronic background events. 
Images showing a total amplitude of more than 40 photoelectrons were included 
in the analysis. 

\subsection{Distributions\label{distributions}}

\begin{figure}[htb]
\begin{center}
\begin{tabular}{cc}
 \mbox{
    \epsfxsize7.0cm
    \epsffile{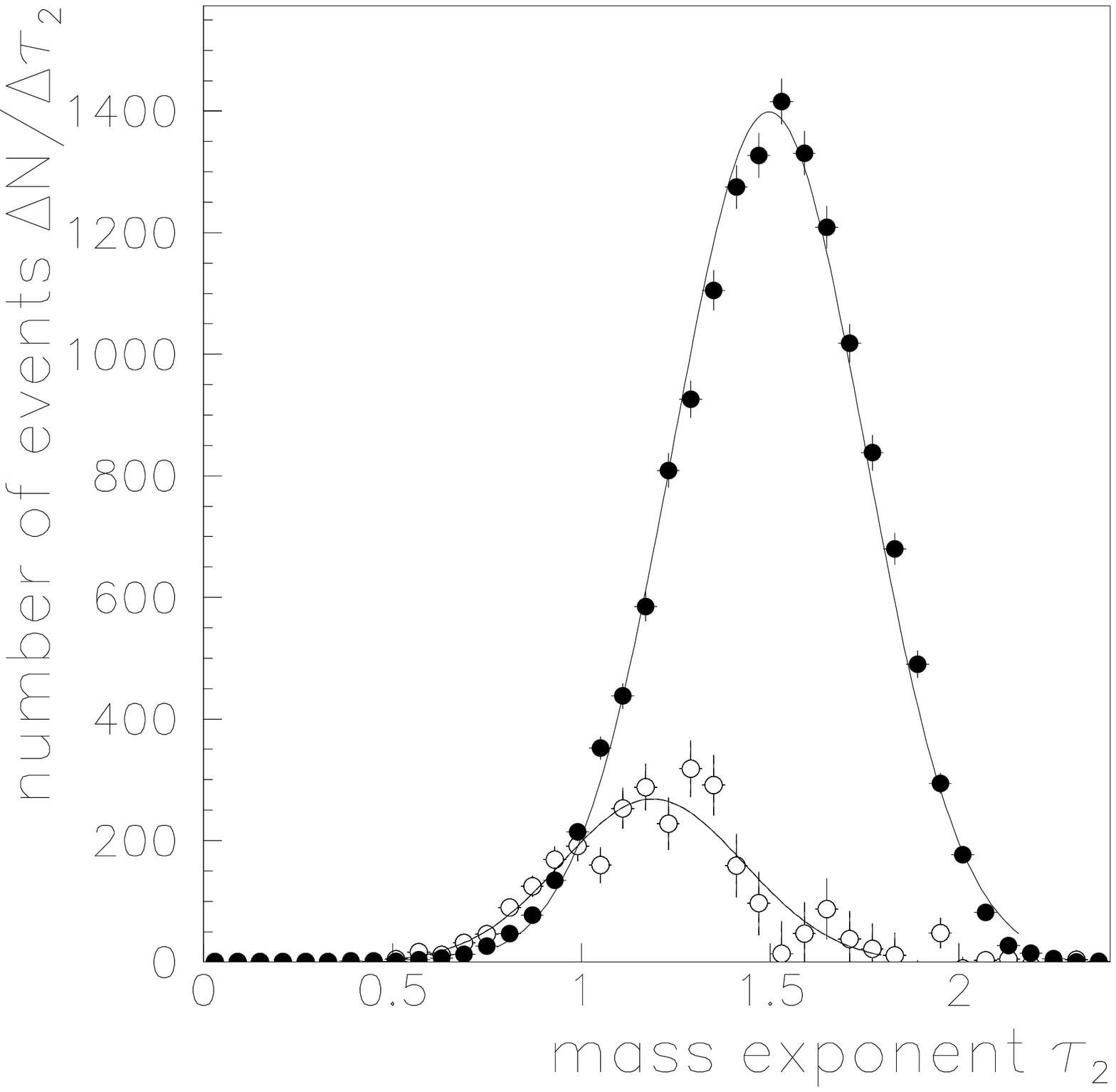}
      }
 & 
 \mbox{
    \epsfxsize7.0cm
    \epsffile{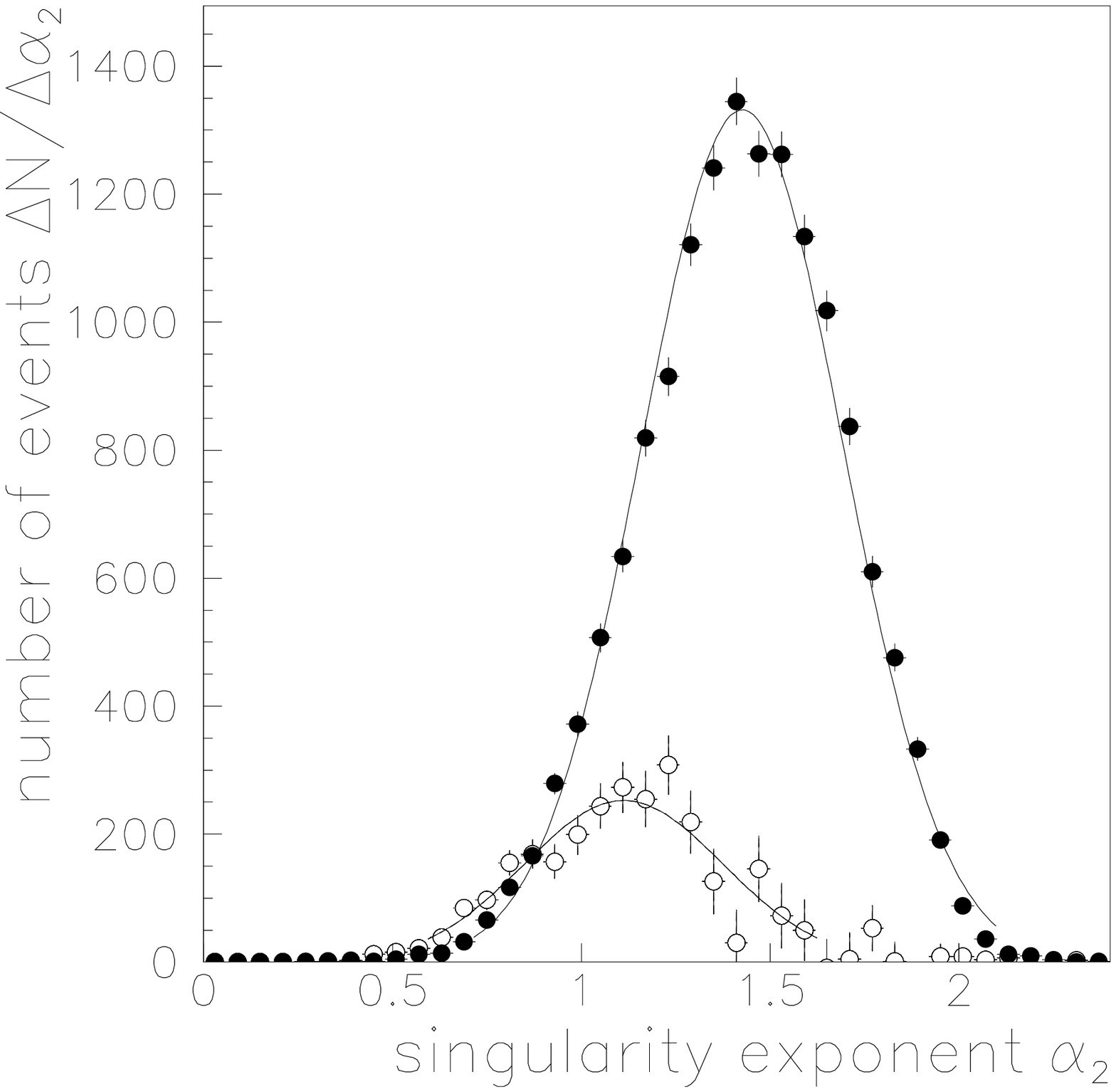}
      }
\end{tabular}
\end{center}
\caption{The distributions of the mass exponent $\tau_2$ (left) and the
singularity exponent $\alpha_2$ (right) for  $\gamma$-rays (open circles) und 
hadrons (filled circles).}
\label{multifracspectrum}
\end{figure}

Figure \ref{multifracspectrum} shows the distributions of the
multifractal parameters $\tau_2$ and $\alpha_2$ for $\gamma$-rays as well 
as for cosmic rays. The distribution is very well described by a Gaussian
function. The multifractal parameter 
$\tau$ indicates how fast the number of cells of pixels that carry a 
significant amount of the total amplitude increases, if the size of the cell 
is decreased. Images originating from a $\gamma$-ray yield smaller 
values for $\tau$, because these images are well contained in a narrow region of 
the camera in contrast to the fuzzy images produced by hadronic showers. 
For these hadronic images, one observes a fast 
increase in the number of cells with decreasing cell size,
and therefore the $\tau$-distribution peaks at larger 
values.\\

The parameter $\alpha$ indicates how fast $\tau$ changes with increasing order
$q$. Considering the images of a hadronic shower, they are
spread out over a large region of the camera and thus the different cells of
pixels carry only small amplitudes, i.e. numbers close to zero in contrast to a
$\gamma$-induced shower, which generates amplitudes that are relatively large in
the various cells of pixels. With increasing q the moments $Z_q(\ell)$ decrease
rapidly in the case of images produced by hadrons, resulting in a fast increase
in the slope $\tau(q)$ with growing order $q$. \\

\begin{figure}[htb]
\begin{center}
 \mbox{
    \epsfxsize7.0cm
    \epsffile{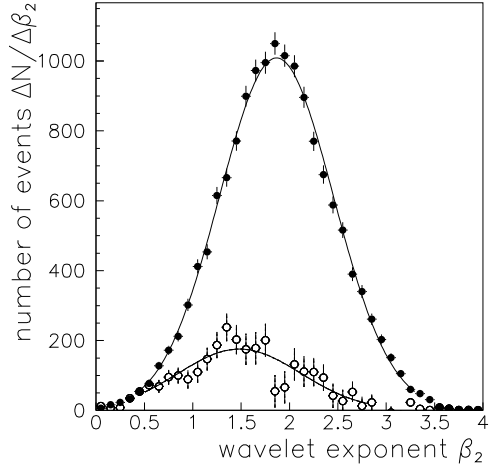}
      }
\end{center}
\caption{The distribution of the wavelet parameter $\beta_2$ for $\gamma$-rays 
(open circles) and hadrons (filled circles).}
\label{waveletspectrum}
\end{figure}

The distribution of the wavelet parameter $\beta_2$ is shown in Figure
\ref{waveletspectrum}. Images of hadronic origin show larger fluctuations on the
short length scales compared to an image produced by $\gamma$-rays, that may in 
fact be described by a smooth gaussian distribution. This results in an increase 
of the wavelet moments $W_q(\ell)$ with decreasing length scale. Therefore, the
parameter $\beta$ assumes larger values for hadronic events compared to
$\gamma$-rays. \\

\subsection{Separation Properties Compared with the Hillas-Parameters}

In order to describe the effectiveness of the different parameters with respect 
to distinguishing between $\gamma$-rays and the hadronic background, cuts on these
parameters were applied. From that, efficiencies $\kappa$ and $Q$-values 
were derived:
$\kappa_{\gamma}=N_{\gamma}^{\mbox{\it\scriptsize (cut)}}/N_{\gamma}$, 
$\kappa_{\mbox{\scriptsize{Hadron}}}=
N_{\mbox{\scriptsize{Hadron}}}^{\mbox{\it\scriptsize (cut)}}/N_{\mbox{\scriptsize{Hadron}}}$ 
and $Q=\kappa_{\gamma}/\sqrt{\kappa_{\mbox{\scriptsize{Hadron}}}}$.
The number $N_{\gamma}$ was determined by statistically subtracting the 
numbers of the events in the source region and in the background region.
The $Q$-value describes by how much the significance of a weak signal, calculated as in \cite{LI83}, is enhanced by the cuts. 
The analysis was first applied to all telescopes individually.\\

Table \ref{comparison} lists mean values and standard
deviations of the distribution for conventional and multifractal image parameters 
for the $\gamma$-rays as well as for the hadron-distributions, 
the $\gamma$-efficiencies $\kappa_{\gamma}$, hadron-efficiencies 
$\kappa_{\mbox{\scriptsize{Hadron}}}$ and $Q$-values for a cut that yields the 
largest $Q$-value in descending order with respect to the $Q$-value they 
archieve. As already evident from Figure~\ref{multifracspectrum}, multifractal 
parameters can be employed to enhance the significance of $\gamma$-ray signals, 
but seem not to be competitive with the conventional parameter 
{\it scaled width}.

\begin{table}[htb]
\begin{center}
\begin{tabular}{|c||c|c|c|c|c|}\hline
& \parbox{1.2cm}{\baselineskip 0.27cm {\it scaled\\width}} 
& $\tau_2$ & $\alpha_2$ 
& \parbox{1.2cm}{\baselineskip 0.27cm {\it scaled\\length}} 
& $\beta_2$ \\ \hline\hline
$\langle x^{\gamma}\rangle$             & 1.04 & 1.18 & 1.12 & 0.99 & 1.45\\
$\sigma^{\gamma}_x$                     & 0.18 & 0.24 & 0.24 & 1.26 & 0.50\\ 
$\langle x^{\mbox{\scriptsize{Hadron}}}\rangle$
                                        & 1.58 & 1.47 & 1.40 & 0.19 & 1.81\\
$\sigma^{\mbox{\scriptsize{Hadron}}}_x$ & 0.38 & 0.24 & 0.26 & 0.29 & 0.56\\ \hline
$\mbox{{\it cut}}^{\mbox{\it{\scriptsize{(opt)}}}}$
                                        & $sw\leq1.1$&$\tau_2\leq1.25$&$\alpha_2\leq1.25$&$sl\leq1.2$&$\beta_2\leq1.9$\\
$\kappa_{\gamma}$                       & $0.663\pm0.044$&$0.722\pm0.050$&$0.797\pm0.056$&$0.710\pm0.006$&$0.879\pm0.067$\\
$\kappa_{\mbox{\scriptsize{Hadron}}}$   & $0.113\pm0.004$&$0.241\pm0.005$&$0.338\pm0.006$&$0.404\pm0.054$&$0.635\pm0.008$\\ \hline
{\it Q-value}                          & $1.82\pm0.12$&$1.47\pm0.19$& $1.37\pm0.09$&$1.12\pm0.08$&$1.10\pm0.08$\\ \hline
\end{tabular}
\caption{Description of the distributions of the various parameters, including 
efficiencies and the $Q$-value for an optimal cut, i.e. for maximum 
enhancement of significance. The values quoted refer to events detected with the 
telescope $\mbox{CT}_3$; values resulting from the analysis of images in the
other telescopes are consistent within statistics. 
}
\label{comparison}
\end{center}
\end{table}

\subsection{Dependence on $q$}

\begin{figure}[htb]
\begin{center}
\begin{tabular}{cc}
 \mbox{
    \epsfxsize7.0cm
    \epsffile{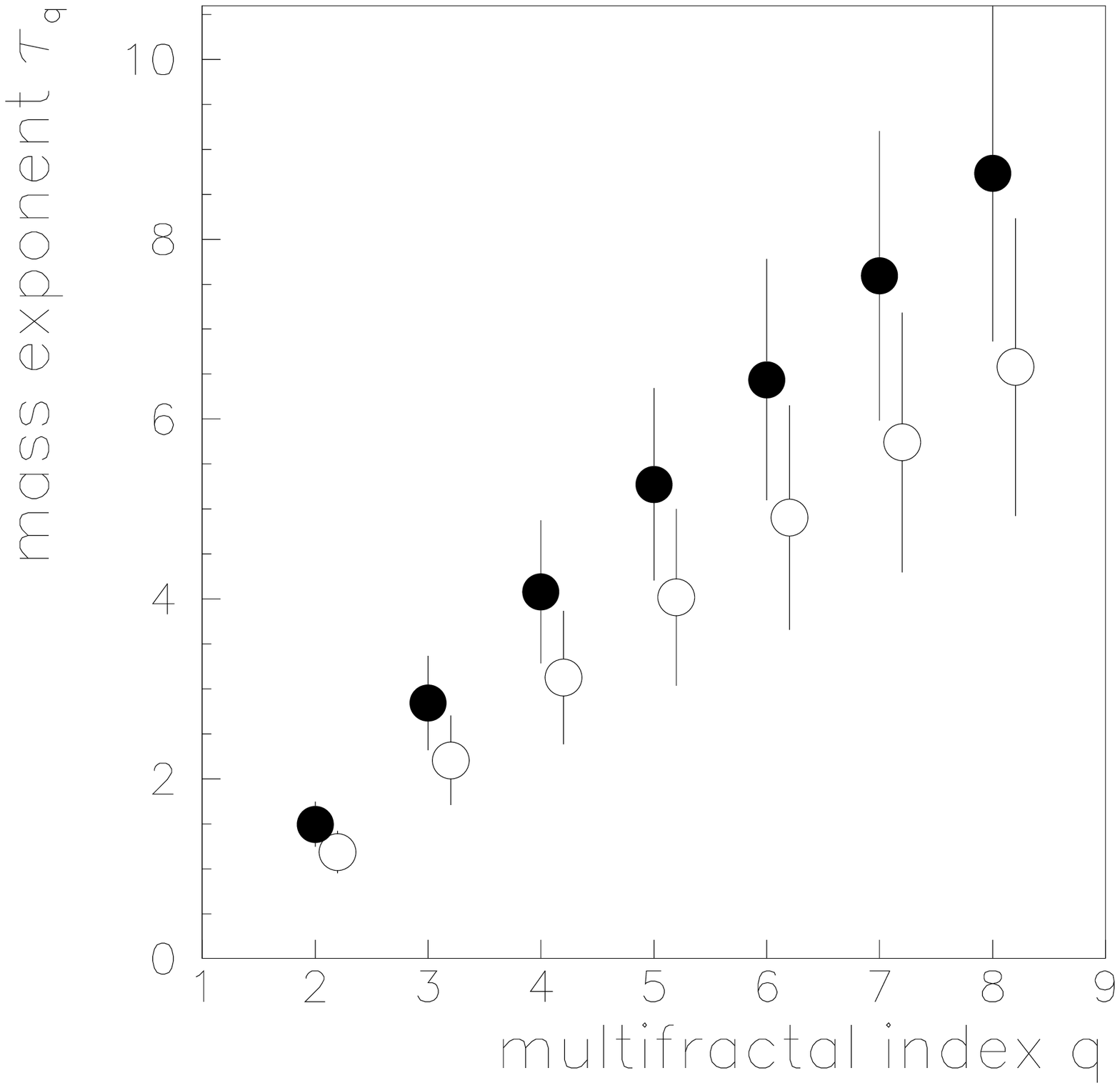}
      }
 & 
 \mbox{
    \epsfxsize7.0cm
    \epsffile{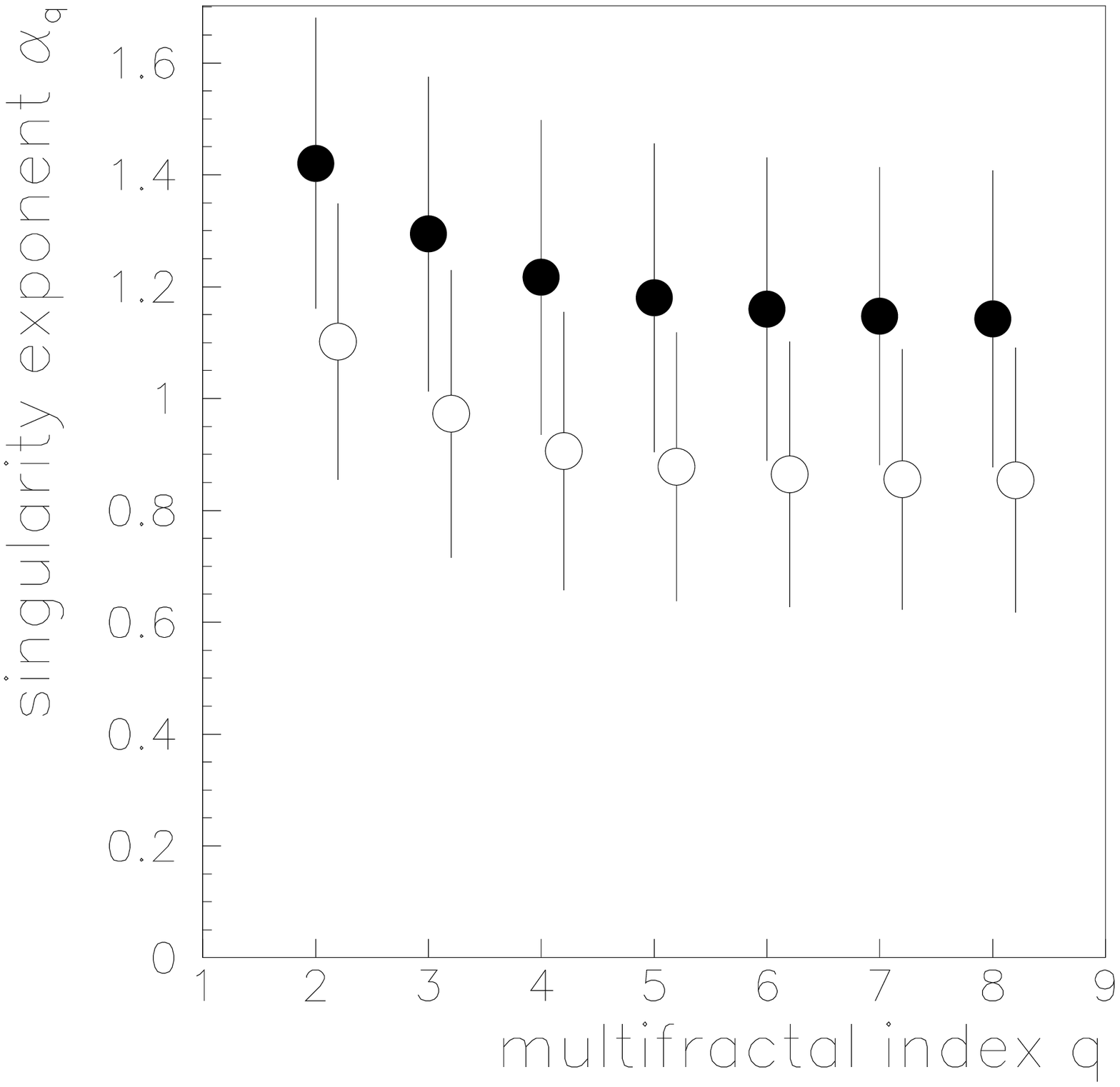}
      }
\end{tabular}
\end{center}
\caption{Mean values and standard deviations of the multifractal parameters
$\tau$ (left) and $\alpha$ (right) as a function of the index $q$ for
$\gamma$-rays (open circles) and hadrons (filled circles).}
\label{qdependency}
\end{figure}

An interesting point is the dependence of the parameters on the index $q$,
especially how the separation performance of the parameters is
influenced by different weighting exponents. Figure \ref{qdependency}
illustrates how the parameters $\tau$ and $\alpha$ change with $q$.
Depicted are the mean
values for the $\gamma$-ray and hadron distributions a well as their widths
resulting from a Gaussian fit.\\

The parameters show the expected dependence on $q$. However, their separation
performance is virtually unaffected by a particular choice of $q$
-- in all cases, the separation between $\gamma$-rays and hadrons
corresponds roughly to the width of the distributions. The reason
is that parameters of different order $q$ turn out to be strongly 
correlated.\\ 

Furthermore, the values of the 
multifractal parameters are affected by the position of the
shower image relative to the pixel cells, clearly an effect of the relatively
large cells.
The value of the moment $Z_q(\ell)$ depends on
the position of the image relative cells of pixels especially for the
largest cell. Also the points in the double-logarithmic plot of 
$Z_q(\ell)$ against $\ell$ do not lie on a line as expected from equation
\ref{loglog}, but rather on a concave curve. For that reason the slope $\tau$ of a 
line through these points gets steeper if one includes only the moments 
$Z_q(\ell)$ on the short length scales. 

\subsection{Correlations}

An interesting question is to which extent the multifractal parameters and
the Hillas parameters are correlated. If multifractal parameters are
genuinely sensitive to fluctations within the image, a weak correlation
would be expected. If, on the other hand, the multifractal parameters 
merely provide a different parametrization of the global shape of
the image, one would expect strong correlations.

\begin{figure}[htb]
\begin{center}
\begin{tabular}{cc}
 \mbox{
    \epsfxsize7.0cm
    \epsffile{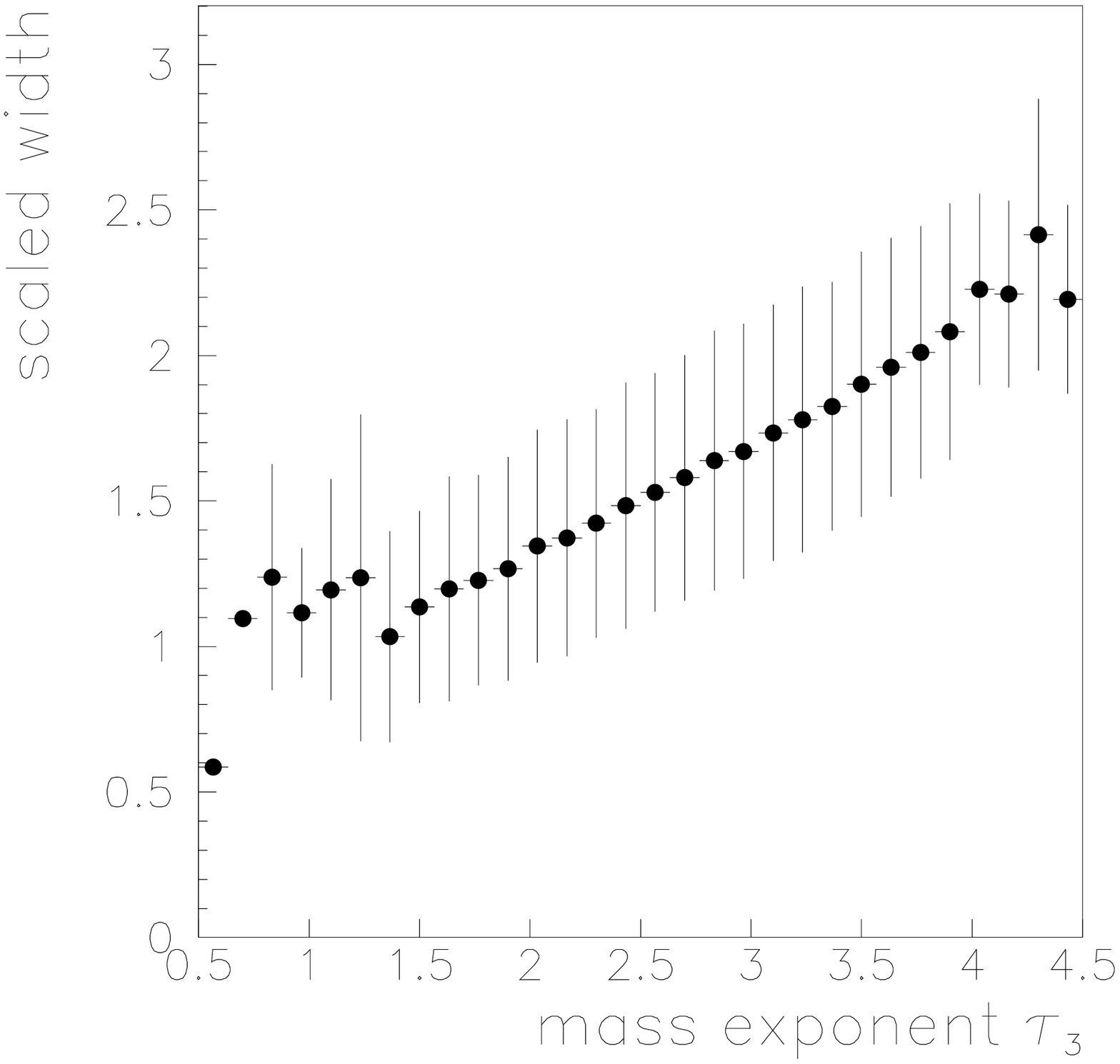}
      } &
 \mbox{
    \epsfxsize7.0cm
    \epsffile{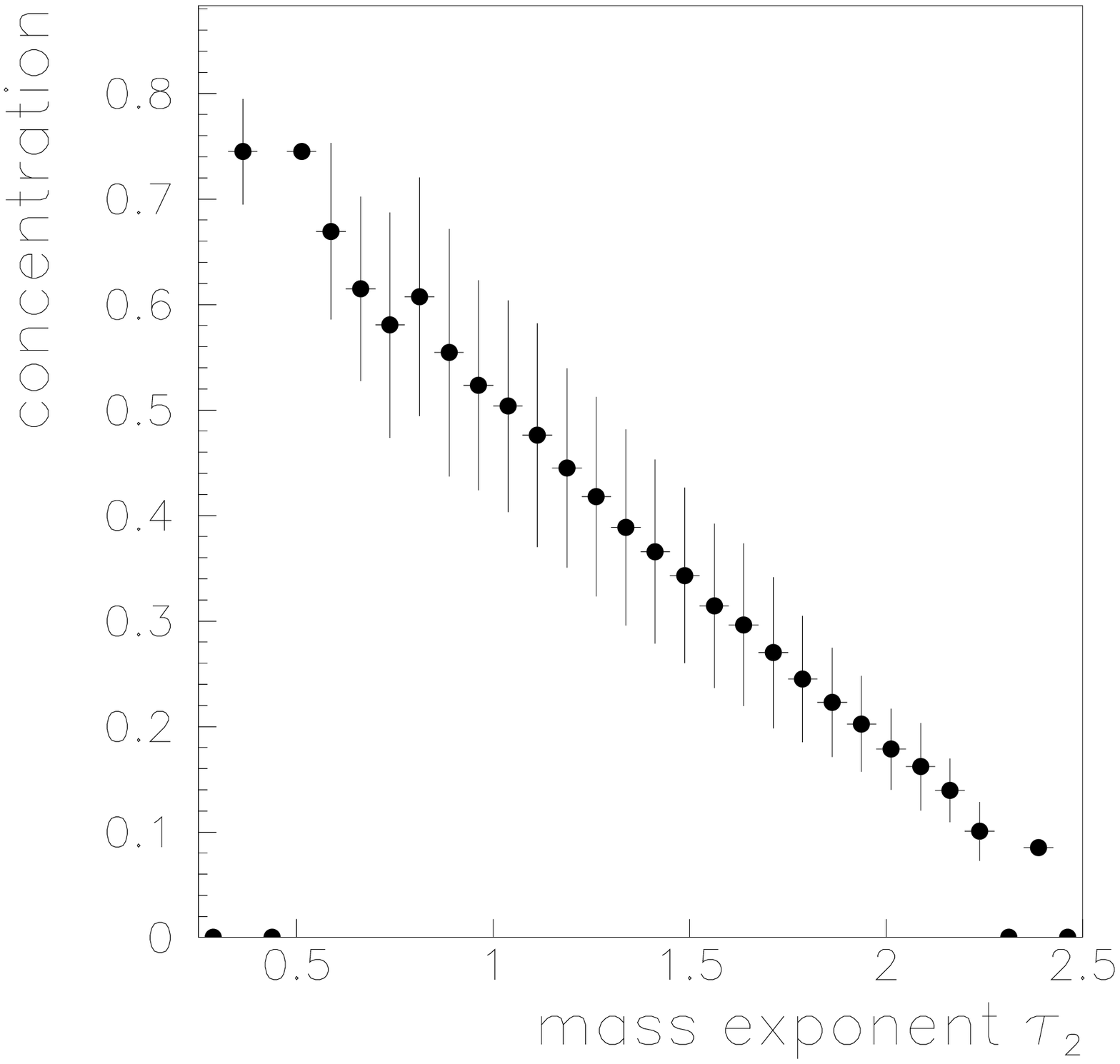}
      } \\
\end{tabular}
\end{center}
\caption{Correlation of the multifractal parameter $\tau$ with the
Hillas-parameters {\it scaled width} (left) and {\it concentration} (right).}
\label{correlation}
\end{figure}

In Figure~\ref{correlation} the correlation between the multifractal
parameter $\tau$ and the Hillas-parameters {\it scaled width} and {\it
concentration} is depicted for the hadronic background. The mean value and the 
standard deviation of the Hillas parameter is plotted as a function of the 
value of a multifractal parameter.\\

Clearly, the parameters are not independent of each other. The correlation of
the multifractal parameter $\tau$ with {\it scaled width} directly follows from
the arguments outlined in section \ref{distributions}. Furthermore, the
correlation of $\tau$ with {\it concentration} shows an accumulation of
$\gamma$-candidates at small values for $\tau$ and large values for {\it
concentration}, whereas hadron-candidates amass at large values for $\tau$ and
small values for {\it concentration}.

\subsection{Averaging over Multiple Telescopes}

In the conventional analysis using the (scaled) Hillas parameters,
separation properties are enhanced by averaging over 
the images taken from the same shower by the different telescopes.
The same is true for the multifractal parameters.
Averaging values of the multifractal parameters $\tau$ or $\alpha$ over all 
telescopes that have been triggered by a given event results in a significant 
improvement of its separation properties, as can be seen in Table \ref{average}.
Still, however, are the multifractal parameters inferior in their performance.

\begin{table}[htb]
\begin{center}
\begin{tabular}{|c||c|c|c|c|}\hline
 &{\it cut}&$\kappa_{\gamma}$&$\kappa_{\mbox{\scriptsize{Hadron}}}$ &$Q$-value\\ \hline
$\tau_2$                & $\tau_2\leq1.17$        &    $0.555\pm0.035$ & $0.076\pm0.003$ & $2.02\pm0.13$ \\
$\beta_2$               & $\beta_2\leq1.60$       &    $0.650\pm0.046$ & $0.318\pm0.005$ & $1.15\pm0.08$ \\ \hline
\mbox{\it mean scaled width}  & $\mbox{\it msw}\leq1.10$ & $0.718\pm0.043$ & $0.074\pm0.003$ & $2.64\pm0.16$ \\
\mbox{\it mean scaled length} & $\mbox{\it msl}\leq1.10$ & $0.724\pm0.039$ & $0.326\pm0.005$ & $1.27\pm0.09$ \\ \hline
\end{tabular}
\end{center}
\caption{Efficiencies and $Q$-values for an optimal cut for the averaged
parameters.}
\label{average}
\end{table}

% +++++++++++++++++++++++++++++++++++++++++++++++++++++++++++++++++++++++++++ %
% +++++                      NEURAL NETWORKS                            +++++ %
% +++++++++++++++++++++++++++++++++++++++++++++++++++++++++++++++++++++++++++ %

\section{Neural Networks}

In order to investigate how the $\gamma$/hadron-separation may be improved 
in spite of the close correlation of the parameters, a neural network was
employed. A detailed introduction to the functionality and theory of neural
nets may be found in \cite{BEA92}. The analysis was applied
to data from individual telescopes. Neural networks of the type 
\mbox{`feed-forward perceptron'} having

\begin{enumerate}
\item{the Hillas-parameters {\it width}, {\it length}, {\it concentration} and
$\log(\mbox{\it size})$;}
\item{multifractal and wavelet parameter $\tau_2$, $\tau_4$, $\tau_8$,
$\beta_2$, $\beta_4$, $\beta_8$ and $\log(\mbox{\it size})$;}
\item{all of the above parameters}
\end{enumerate}

as input values were implemented and their classification performance was
examined. The adjustment of the net's weights and thresholds was done using a
data sample consisting of 1200 events with a reconstructed shower direction
within a circle of $0.2^{\circ}$ around the source-position as $\gamma$-rays and 
1200 events of background data as hadrons. The events were requested to have an
image-size of at least 40 photoelectrons. The network was adjusted in such a 
way that it produced an output value of $z=0$ for the $\gamma$-candidates and
$z=1$ for the hadron-candidates, respectively.\\

The neural nets were implemented using the {\it JetNet}-programme
(\cite{{LOE91}}), that provides a number of interesting features
for optimizing the adjusting process such as adding momentum or introducing 
artificial noise, in order to prevent the system from being trapped in a local 
minimum. The topology of the network was chosen in such a way that the networks
consisted of three layers in all three cases, but the number of nodes within a
layer was altered until the net yielded the best result on the training data.

\begin{figure}[htb]
\begin{center}
 \mbox{
    \epsfxsize7.0cm
    \epsffile{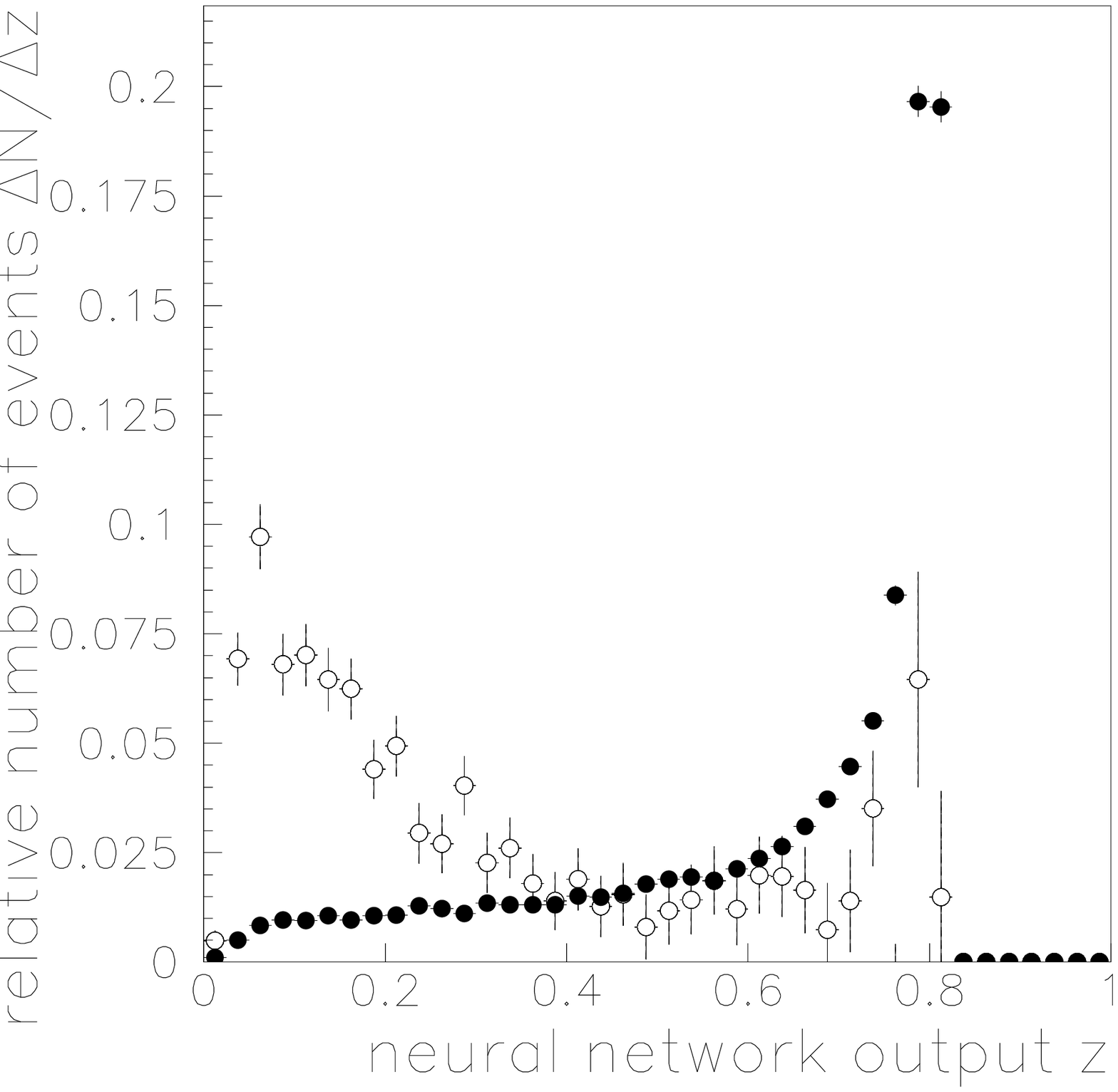}
      }
\end{center}
\caption{The distribution of the output $z$ of the neural net using Hillas- as
well as multifractal parameters; filled circles correspond to the hadronic 
background, open circles to $\gamma$-rays. Both distributions have been
normalized to unit area.}
\label{neuralnet}
\end{figure}

Figure \ref{neuralnet} illustrates the distribution of the output $z$ of the 
neural network, clearly showing an excess of events with low values for $z$
corresponding to $\gamma$-rays.

\begin{table}[htb]
\begin{center}
\begin{tabular}{|c||c|c|c|c|}\hline
 & {\it cut} & $\kappa_{\gamma}$ & $\kappa_{\mbox{\scriptsize Hadron}}$ &
 $Q$-value\\ \hline\hline
1. & $z\leq0.38$ & $0.70\pm0.05$ & $0.15\pm0.004$ & $1.79\pm0.12$ \\
2. & $z\leq0.34$ & $0.54\pm0.04$ & $0.12\pm0.003$ & $1.59\pm0.11$ \\
3. & $z\leq0.26$ & $0.54\pm0.05$ & $0.07\pm0.003$ & $2.01\pm0.14$ \\ \hline
\end{tabular}
\end{center}
\caption{Efficiencies and Q-values for optimal cuts on the neural net output
$z$, comparing the various approaches.}
\label{neuraltable}
\end{table}

The network based on Hillas-parameters was able to reach the performance of the
parameter {\it scaled width}, that attains a $Q$-value of $1.82\pm0.12$.
The network using all available parameters slightly surpasses the performance of the
network that relied on Hillas-parameters. The network based on 
multifractal and wavelet parameters falls behind, as can be seen from Table
\ref{neuraltable}.\\

% +++++++++++++++++++++++++++++++++++++++++++++++++++++++++++++++++++++++++++ %
% +++++                       CONCLUSIONS                               +++++ %
% +++++++++++++++++++++++++++++++++++++++++++++++++++++++++++++++++++++++++++ %

\section{Concluding Remarks}

Image analysis techniques based on multifractals and wavelets were 
applied to images recorded by the HEGRA-{\CE}-telescopes.
A $\gamma$/hadron-separation by cuts on these parameters is
feasible, yielding a performance similar to the Hillas-parameter
{\it concentration}, but falling behind that of the parameter {\it scaled
width}. Combining Hillas parameters and multifractal parameters using a neural
network, a slight improvement in performance could be achieved compared
to the Hillas parameters alone. The gain is, however, limited by the 
strong correlation between the Hillas parameters and the multifractal 
parameters.

A general -- and not too surprising -- conclusion is certainly that
the pixel size of the HEGRA cameras is much too coarse to effectively
characterize the fractal nature of showers images; rather than
exploring the fine structure of the image, the multifractal parameters
determine its overall shape, explaining the good correlation
with the Hillas parameter {\it concentration}.

Our overall assessment concerning the use of multifractal
parameters to identifiy the primary particle of an air shower
is less optimistic than the view given in (\cite{HAU99}), in which the analysis
method is applied to Monte-Carlo data from a simulation of the TACTIC-experiment.
While the pixel size of HEGRA ($0.25^\circ$) and TACTIC ($0.31^\circ$)
are similar, some basic differences between the two analysis should
be emphasized:
\begin{itemize}
\item The square-grid pixel arrangement of TACTIC simplifies the subdivision
into pixel cells.
\item The analysis of \cite{HAU99} is applied to images containing at
least 1800 photoelectrons, whereas the HEGRA analysis was applied to
images with as little as 40 photoelectrons. For such small numbers of
photoelectrons, fluctuations in the image are dominated by photoelectron
statistics, rather than the shower substructures. For gamma-hadron
separation, algorithms which work only for very large images are of
limited interest. The situation differs if one applies the technique
to separate primary nuclei, which is the main emphasis of \cite{HAU99}.
\item The analysis of \cite{HAU99} also included multifractal moments
for negative exponents $q$. This was possible since in the simulation
Poisson-distributed background photoelectrons were added and there is
a clear distinction between 0 and 1 photoelectron in a pixel. In HEGRA,
with a Gaussian noise of about 1 photoelectron rms, an ad-hoc cut would
have to be used to exclude low amplitudes; to avoid this, only positive
values of $q$ were allowed.
\end{itemize}

In order to really judge if one can detected the 
fractal structure of air showers
in their {\CE} images, one should probably employ a simulation with very
fine pixels -- such that the first set of length scales is well contained 
within the image -- and a photoelectron yield such that Poisson fluctuations
are small compared to the shower fluctuations. The first condition
implies a pixel size around $0.002^\circ$ to $0.01^\circ$, the second
an image {\it size} of 10000 to 100000 photoelectrons.

\section*{Acknowledgements}

The support of the HEGRA experiment by the German Ministry for Research 
and Technology BMBF and by the Spanish Research Council
CYCIT is acknowledged. We are grateful to the Instituto
de Astrof\'{\i}sica de Canarias for the use of the site and
for providing excellent working conditions. We thank the other
members of the HEGRA CT group, who participated in the construction,
installation, and operation of the telescopes. We gratefully
acknowledge the technical support staff of Heidelberg,
Kiel, Munich, and Yerevan.

% -- bibliography -- %

\end{document}